\documentclass[prb,aps,twocolumn,showpacs,nofootinbib]{revtex4}
\usepackage{graphicx}
\usepackage{amsmath}
\usepackage{epstopdf}
\usepackage{amsbsy}
\usepackage{color}

\usepackage{bm} 

\begin{document}

\title{Bulk and Surface Nanoscale Hole Density Inhomogeneity in 
HgBa$_2$CuO$_{4+\delta}$ and Bi$_2$Sr$_2$CaCu$_2$O$_{8+\delta}$ Cuprates}
\author{Wei Chen$^{1}$, Giniyat Khaliullin$^{2}$, and Oleg P. Sushkov$^{1}$}
\affiliation{$^{1}$School of Physics, University of New South Wales, 
Sydney 2052, Australia \\
$^{2}$Max-Planck-Institut f\"{u}r Festk\"{o}rperforschung, 
Heisenbergstrasse 1, D-70569 Stuttgart, Germany}

\date{\today}

\begin{abstract}
It is well established that the hole density in the prototypical 
superconductor La$_{2-x}$Sr$_x$CuO$_4$ is very inhomogeneous due to Sr-dopant 
induced disorder. On the other hand, the hole distribution in 
HgBa$_2$CuO$_{4+\delta}$ and Bi$_2$Sr$_2$CaCu$_2$O$_{8+\delta}$ doped by 
interstitial oxygen is believed to be much more uniform. Recent nuclear 
magnetic resonance measurements indicate, however, that the charge 
inhomogeneity in HgBa$_2$CuO$_{4+\delta}$ is close to that in 
La$_{2-x}$Sr$_x$CuO$_4$. Calculations performed in the present paper confirm 
this observation. We also show that the charge inhomogeneity is most 
pronounced at the surface layer that can be probed by scanning tunneling 
microscope. Our simulations demonstrate that, despite having similar 
amplitudes of charge inhomogeneity, the hole mean free path in 
HgBa$_2$CuO$_{4+\delta}$ is substantially longer than that in 
La$_{2-x}$Sr$_x$CuO$_4$. The screening of the Coulomb repulsion in 
HgBa$_2$CuO$_{4+\delta}$ is also stronger. These two reasons might explain 
the difference in the superconducting critical temperatures between these 
two compounds. 
\end{abstract}

\maketitle

\section{Introduction}
The nanoscale electronic disorder is a long standing problem in physics
of cuprates. This problem has many aspects, among which  
the most important one is the influence of disorder on critical 
temperature $T_{c}$. Since the energy scale associated with pairing mechanism 
in CuO$_{2}$ plane is believed to be universal, different out-of-plane defects 
are expected to influence $T_{c}$ differently.\cite{Eis04,Rul08} 
Another important aspect is the nanoscale inhomogeneity of the local 
density of states (DOS) measured by scanning tunneling microscope (STM).
This effect was mostly studied~\cite{McElroy05,Gomes07,Pasupathy08} in 
Bi$_2$Sr$_2$CaCu$_2$O$_{8+\delta}$. In this paper, we address theoretically the 
problem of the nanoscale charge inhomogeneity in cuprates, 
and a related issue of the hole mean free path. We focus on two single layer 
families of high-$T_c$ superconductors, La$_{2-x}$Sr$_{x}$CuO$_{4}$ and 
HgBa$_2$CuO$_{4+\delta}$, in which the density inhomogeneity has been quantified 
by nuclear quadrupole resonance (NQR) experiments.~\cite{Singer02,Rybicki09}
We also calculate the charge inhomogeneity in the surface layer of 
Bi$_2$Sr$_2$CaCu$_2$O$_{8+\delta}$.

The NQR measures the energy splitting of nuclear levels induced by an 
electric field gradient at the nucleus. A hole in the 3d-shell of the Cu ion 
gives a dominant contribution to the field gradient at the Cu nucleus. The 
Cu NQR frequency in cuprates is therefore very sensitive to the doping level 
and is directly proportional to the local hole concentration. The contribution 
of the 3d-hole to the field gradient is significantly compensated by 
contributions of holes located at nearby oxygens,~\cite{Ohta,Mori}
therefore the slope of the NQR frequency versus doping varies among 
different cuprate families. 

It is widely accepted that the hole density distribution within the CuO$_2$ 
layer in La$_{2-x}$Sr$_x$CuO$_4$ is very nonuniform, as it has been clearly 
demonstrated by measurements of $^{63}$Cu NQR spectra~\cite{Singer02}. 
The observed broad NQR spectrum~\cite{Singer02} unambiguously indicates a 
very inhomogeneous hole density profile in the bulk of the sample. The 
inhomogeneity is due to the doping mechanism, where Sr substitutions of La 
ions create an effective Coulomb defect very close to the conducting CuO$_2$ 
plane. In a recent paper,~\cite{Chen09} we have performed the Hartree-Fock 
simulation of the charge density distribution in La$_{2-x}$Sr$_x$CuO$_4$, 
which shows a very inhomogeneous charge density profile at the nanometer 
scale, and reproduces the observed NQR lineshapes.  

On the other hand, it is generally believed that hole density distribution in 
compounds doped by interstitial oxygens, such as HgBa$_2$CuO$_{4+\delta}$ and 
Bi$_2$Sr$_2$CaCu$_2$O$_{8+\delta}$, is much more uniform than in 
La$_{2-x}$Sr$_x$CuO$_4$. This is because the distance from the interstitial 
oxygen to the CuO$_2$ layer is typically larger than the Sr-layer distance in 
La$_{2-x}$Sr$_x$CuO$_4$. However, Cu NQR measurements in the  
single layer HgBa$_2$CuO$_{4+\delta}$~\cite{Gippius97,Rybicki09} as well as in 
the bilayer HgBa$_2$CaCu$_2$O$_{6+\delta}$~\cite{Jul96} show fairly large 
linewidths comparable to that in La$_{2-x}$Sr$_x$CuO$_4$. As it will 
be demonstrated below, the NQR data imply the same degree of the hole 
density inhomogeneity in La$_{2-x}$Sr$_x$CuO$_4$ and HgBa$_2$CuO$_{4+\delta}$. 
We will also show that despite having similar amplitudes of the charge 
inhomogeneity, the spatial profiles of the density distribution are very 
different in these two cases: it is much smoother in the oxygen doped 
HgBa$_2$CuO$_{4+\delta}$. Consequently, the forward scattering is predominant 
and the mean free path in HgBa$_2$CuO$_{4+\delta}$ is substantially longer
than that in La$_{2-x}$Sr$_x$CuO$_4$. Our simulation also indicates that 
the screening of the Coulomb repulsion between the charge carriers in 
HgBa$_2$CuO$_{4+\delta}$ is stronger. We argue that these two reasons might 
explain the difference in superconducting critical temperatures between 
these two compounds.

The nanoscale charge density inhomogeneity in underdoped cuprates is an
indirect way to distinguish the large Fermi surface of a normal Fermi liquid 
and the small Fermi surface of a doped Mott insulator. Calculations in 
Ref.~\onlinecite{Chen09} for La$_{2-x}$Sr$_x$CuO$_4$ were based on the small 
hole pocket Fermi surface, which implies a small number of mobile charge 
carries. The small number of charge carries results in poor 
screening~\cite{Chen09,Wan02} and hence in the strong charge density 
inhomogeneity consistent with NQR data. On the other hand, the large Fermi 
surface implies the large number of mobile charge carriers and a very 
effective Coulomb screening. In this case calculations give a very moderate 
charge inhomogeneity~\cite{Nunner05,com1}, which is too weak to explain 
the observed NQR linewidths.

Another important point is that the charge inhomogeneity in the surface 
layer is always stronger than that in the bulk layer. This is because the 
surface ionic dielectric constant is about two times smaller than that in 
the bulk~\cite{BT}. Hence the screening at the surface is weaker and this 
results in the stronger charge inhomogeneity.

In this paper, we present Hartree-Fock calculations in underdoped cuprates 
with realistic parameters to simulate the bulk charge distributions in 
HgBa$_2$CuO$_{4+\delta}$ and La$_{2-x}$Sr$_x$CuO$_4$, and the surface charge 
distribution in Bi$_2$Sr$_2$CaCu$_2$O$_{8+\delta}$. Our calculation covers 
from underdoped to optimally doped regime. We fine tune theoretical 
parameters to reproduce experimentally known NQR 
linewidths~\cite{Singer02,Rybicki09}. This enables us to perform a very 
accurate comparison of HgBa$_2$CuO$_{4+\delta}$ and La$_{2-x}$Sr$_x$CuO$_4$. 
We find the charge inhomogeneity of a similar scale in both cases. However, 
the landscapes of spatial modulations and the hole mean free paths are 
substantially different. This is due to different positions of the dopant 
oxygen and the Sr-ion relative to the CuO$_2$ plane. In addition, the 
dielectric constants in these two compounds are different.

Motivated by STM data~\cite{McElroy05,Gomes07,Pasupathy08} in 
Bi$_2$Sr$_2$CaCu$_2$O$_{8+\delta}$, we calculate also the hole density
distribution in the surface CuO$_2$ layer of this compound. The results 
obtained show a large charge density inhomogeneity comparable to that in the 
bulk of HgBa$_2$CuO$_{4+\delta}$. Naturally, the local charge density is 
highly correlated with the interstitial oxygen positions as has been
noticed previously~\cite{Wan02}. 

Structure of the paper is the following: in Sec. II we formulate the
effective model for an isolated CuO$_2$ layer. Because of the long-range 
nature of the Coulomb  interaction, however, the isolated layer approximation 
is not sufficient and one has to take into account other layers. The
impact of the other layers depends on the way of doping and on structural 
details: whether this is a single layer or a double layer compound. The single 
layer La$_{2-x}$Sr$_x$CuO$_4$ and HgBa$_2$CuO$_{4+\delta}$ are considered in 
Section III. In Sec. IV we simulate the charge distribution on the surface of 
the double layer Bi$_2$Sr$_2$CaCu$_2$O$_{8+\delta}$. We also calculate the 
correlation function between the local charge density and oxygen dopant 
positions, and compare the result with the local DOS correlation function 
measured by STM. Our conclusions are presented in Sec V.

\section{The effective model for a $\mbox{CuO}_2$ plane} 

We adopt the effective model formulated in Ref.~\onlinecite{Chen09} based on 
the picture of a lightly doped Mott insulator. Throughout the paper we denote 
the average hole concentration by $p$ and assume $p \ll 1$. The central point 
is that the number of charge carriers is $p$ instead of $1-p$ as one would 
expect for a normal Fermi liquid. We consider first an ``isolated'' CuO$_2$ 
layer. The ``isolated'' means that we disregard screening effects 
of the other layers.

To construct a model relevant to cuprates, we first notice that there are the 
following distinct length scales: ({\it i}) The scale of the order of 1-2 
lattice spacing where the doped holes are dressed by multiple virtual
magnons. ({\it ii}) The scale about average separation between 
Coulomb defects or average separation between holes $\sim 1/\sqrt{p}$. This 
is the scale of the Coulomb screening. ({\it iii}) The scale 
$ r \gg 1/\sqrt{p}$. The Coulomb gap may develop at this scale due to 
Anderson localization effects. 

\begin{figure}[h]
\centering
\includegraphics[width=0.5\columnwidth,clip=true]{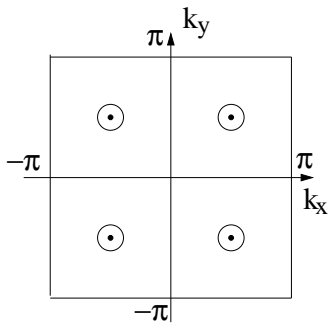}
\caption{Dispersion minima of the spinless fermion generated by Hamiltonian 
(\ref{ht}).}
\label{fig:Disp}
\end{figure}

Regarding the first point, we do not treat the strong correlations 
explicitly, but instead adopt the effective hole dispersion after quantum 
fluctuations at short distances are included. To stress this point, we
frequently use the term ``holon'' instead of ``hole''. It is 
known that the holon dispersion has minima at momenta  
${\bf k}_0=(\pm\pi/2,\pm\pi/2)$, and is approximately isotropic around these 
points~\cite{sushkov97}. The bandwidth of the holon is about $2J$, where 
$J\approx 130$ meV is the superexchange in the $t$-$J$ model, although we 
do not directly employ the $t$-$J$ formalism. Hereafter we set the following 
energy and distance units 
\begin{eqnarray}
\label{j}
J&=&130 \ meV \to 1 \;, \nonumber\\
a_{0}&=&0.38nm\rightarrow 1 \;,
\end{eqnarray}
where $a_0$ is the lattice spacing of the CuO$_2$ plane. To imitate the 
holon dispersion we consider spinless fermions on a 2D square lattice. 
The Hamiltonian reads as follows: 
\begin{equation}
\label{ht}
H_{t}=\sum_{\langle ij \rangle}t^{\prime\prime} c^{\dag}_{i}c_{j}\;,
\end{equation}  
where $c_{i}^{\dag}$ is the holon creation operator at site $i$, and 
$t^{\prime\prime}$ denotes the next-next-nearest-neighbor hopping on the 
square lattice. The Hamiltonian (\ref{ht}) yields the following dispersion 
\begin{eqnarray}
\label{ht1}
\epsilon_{\bf k}=2t^{\prime\prime}(\cos 2k_{x}+\cos 2k_{y})\;.
\end{eqnarray}
The dispersion is isotropic around minima at ${\bf k}_0=(\pm\pi/2,\pm\pi/2)$ as 
shown in Fig.~\ref{fig:Disp}.
We choose $t^{\prime\prime}=0.25J$ to reproduce the realistic holon 
band width, about $2J$, as obtained from numerical simulations of the 
$t-J$ model.~\cite{sushkov97} An additional argument supporting this value 
is as follows. Near its minimum the dispersion (\ref{ht}) can be expanded as
$\epsilon_{\bf k}=\to const +4t^{\prime\prime}|{\bf k}-{\bf k}_0|^2$,
so the holon effective mass is equal to $m^*=\hbar^2/(8a_0^2t^{\prime\prime})$.
The value $t^{\prime\prime}=0.25J$ results in the effective mass 
$m^* \simeq 2m_e$ which is close to the effective mass measured in magnetic 
quantum oscillation experiments~\cite{Sebastian09_2,Singleton09}.
The realistic holon band width and the realistic effective mass justify our 
choice of $t^{\prime\prime}=0.25J$. We note also that in the original 
$t$-$J$ model formalism, there are four holon half-pockets inside magnetic 
Brillouin zone, and each pocket has two pseudospins~\cite{sushkov97}; 
in the present model, we consider four full pockets inside the full Brillouin 
zone with spinless fermions, hence the number of {\it charge} degrees of 
freedom is exactly the same.~\cite{noteMBZ} 

Even though the holes are heavily dressed by magnetic fluctuations, their 
charge is conserved and hence they interact with Coulomb defects via the  
ordinary Coulomb potential: 
\begin{eqnarray}
\label{HUQ}
H_{h-O}&=&\sum_{l,i}{\cal U}_{li}c_{i}^{\dag}c_{i} \;, \nonumber\\
{\cal U}_{li}&=&\frac{-Q}{\sqrt{|{\bf R}_{l}-{\bf r}_{i}|^{2}+a_d
^{2}}} \;, \nonumber\\
a_d^2&=&a_{ZR}^2+\lambda^2 \;.
\end{eqnarray}
Here ${\bm r}_i$ is position of the holon and ${\bm R}_l$ is the in-plane
projection of the Coulomb defect (Sr-ion or dopant oxygen) position. The 
distance from the plane to the defect is $\lambda$, and $a_{ZR}\approx 0.8$ 
is the size of the Zhang-Rice singlet. (We recall that the energy and distances 
are given in units of $J$ and $a_0$, correspondingly). The dimensionless 
''charge'' $Q \sim 0.5$ depends on the compound, and we discuss its precise 
values later.

Holon-holon Coulomb interaction is of a similar form
\begin{eqnarray}
\label{Hhh}
H_{int}&=&\sum_{ij}U_{ij}c^{\dag}_{i}c_{i}c^{\dag}_{j}c_{j}  \;, \nonumber\\
U_{ij}&=& \frac{Q}{\sqrt{|{\bf r}_{i}-{\bf r}_{j}|^{2}+a_{hh}^{2}}} \ ,
\end{eqnarray}
where $a_{hh}^2\approx 2a_{ZR}^2\approx 1$ stands for the combined size of two 
Zhang-Rice singlets.

The total Hamiltonian
\begin{eqnarray}
H=H_{t}+H_{h-O}+H_{int}\;,
\label{H_total}
\end{eqnarray}
describes the in-plane Coulomb problem. Since the Coulomb interaction is not 
very strong, we solve the many-body problem with the Hamiltonian 
(\ref{H_total}) using the standard Hartree-Fock method. In other words we 
use the Hartree-Fock decomposition of the Coulomb interaction between holons: 
\begin{equation}
\label{hhf}
H_{int}\to
\sum_{ij}U_{ij}\langle c^{\dag}_{i}c_{i}\rangle c^{\dag}_{j}c_{j}
-\sum_{ij}U_{ij}\langle c^{\dag}_{i}c_{j}\rangle c^{\dag}_{j}c_{i}\;.
\end{equation}
This can be done for zero as well as for finite temperatures, 
as it is described in Ref.~\onlinecite{Chen09}. 
 
The above formulation would solve the Coulomb problem for the ``isolated'' 
CuO$_2$ layer.  However, the layer is always embedded in a multilayer 
structure, and because of the long-range nature of the Coulomb interaction, 
we have to take into account other layers. Their influence depends on the 
lattice structure. In the next section, we consider two different families 
of single layer cuprates. 

\section{Single layer $\mbox{HgBa}_{2}\mbox{CuO}_{4+\delta}$
and $\mbox{La}_{2-x}\mbox{Sr}_x\mbox{CuO}_4$ compounds. 
Charge density distribution, NQR line shape,  density of states}

\begin{figure}[htb]
\centering
\leavevmode
\includegraphics[width=0.75\columnwidth,clip=true]{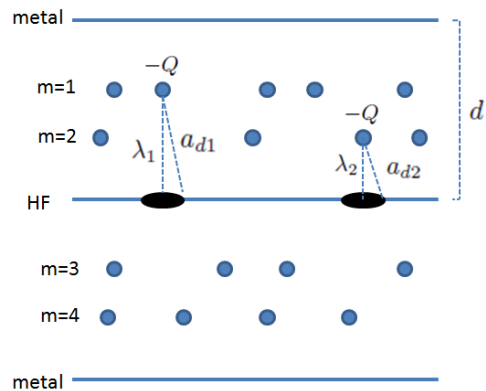}
\caption{(Color online) The Hartree-Fock model of La$_{2-x}$Sr$_x$CuO$_4$
with four ($m=1,2,3,4$) layers of Coulomb defects, i.e., doped Sr-ions, 
and with two ''metallic'' sheets.}
\label{fig:LSCO_structure}
\end{figure}

We treat a particular CuO$_{2}$ plane using the Hartree-Fock (HF) method.
The role of other CuO$_{2}$ planes is to provide screening of the Coulomb 
interaction in the Hartree-Fock analysis. Since CuO$_{2}$ planes have a very 
high longitudinal polarizability we consider the ``other planes'' as purely 
metallic. It has been  shown in Ref.~\onlinecite{Chen09} that this ``metallic 
approximation'' is valid at $p > 1-2\%$ when the polarizability is 
sufficiently high. Within this approximation the HF plane in a single layer 
compound is sandwiched between two ''metallic'' sheets, as demonstrated in 
Fig.~\ref{fig:LSCO_structure} for $\mbox{La}_{2-x}\mbox{Sr}_x\mbox{CuO}_4$.

\subsection{La$_{2-x}$Sr$_x$CuO$_4$}  
Coulomb defects in La$_{2-x}$Sr$_x$CuO$_4$  are created by Sr substitution 
for La ions. Therefore each defect donates one hole
\begin{equation}
p=x \;. 
\end{equation}
Given the periodic structure of CuO$_{2}$ planes along $c-$axis, there are 
two layers of Sr defects between two neighboring CuO$_{2}$ planes, as shown 
in Fig.~\ref{fig:LSCO_structure}. We mark the defect layers by the index 
$m=\left\{1,2,3,4\right\}$, so the HF CuO$_{2}$ plane is under the influence 
of 4 layers of Sr defects. Concentration of defects in each defect layer is 
$x/2$. The planar positions of defects are assumed to be random, and defects 
are assumed to sit above the center of the Cu plaquette. In addition, 
in each defect layer we impose a condition that defects never sit next to 
each other, i.e., the distance between defects is always larger than 
$\sqrt{2}$. The distance between CuO$_2$ layers is $d=1.75$, and the geometric 
distances to Coulomb defects are 
$\lambda_{1}=\lambda_{4}=1.15$, $\lambda_{2}=\lambda_{3}=0.6$.
After accounting for the screening by ``metallic'' planes, the holon 
interaction with a defect (\ref{HUQ}) is replaced by
\begin{eqnarray}
\label{scr}
{\cal U}_{li}^{m}&\to&-Q\left(
\frac{1}{\sqrt{|{\bf R}_{l}-{\bf r}_{i}|^{2}+a_{dm}|
^{2}}}\right. \\
&&+\sum_{n=1}^{\infty}\frac{(-1)^n}
{\sqrt{|{\bf R}_{l}-{\bf r}_{i}|^{2}+(2nd+\lambda_{m})^{2}}}
\nonumber \\
&&\left.+\sum_{n=1}^{\infty}\frac{(-1)^n}
{\sqrt{|{\bf R}_{l}-{\bf r}_{i}|^{2}+(2nd-\lambda_{m})^{2}}}\right) \;.
\nonumber
\end{eqnarray}
The summation over $n$ reflects the image method for metallic 
screening.~\cite{Chen09} With values of the distances $\lambda$ given 
above, the parameter $a_{d}$, defined in Eq.~(\ref{HUQ}), takes the following 
values: $a_{d1}=a_{d4}=1.4$, and $a_{d2}=a_{d3}=1$. The effective charge Q is 
determined by the dielectric constant of the lattice $\epsilon_l$.
In our previous work~\cite{Chen09} we used the value $\epsilon_l=40$
that approximately corresponds to the undoped compound.~\cite{Chen91}
This gave a reasonable fit of the NQR lines, but still the widths were about 
20-30\% larger compared to experiment. In the present paper, we fine tune the 
NQR widths by using $\epsilon_l$ as a fitting parameter. Our best fit yields
\begin{eqnarray}
LSCO: \ \ \ \epsilon_l=30(1+6.25p)\;.
\label{epsilon_LSCO}
\end{eqnarray}
Thus the value of $\epsilon_l$ depends on doping, $\epsilon_l=30$ for the 
undoped compound~\cite{Chen91}, and $\epsilon_l=60$ for the optimally doped 
compound ($p=0.16$). The coefficient 6.25 has been obtained by fitting the 
NQR widths, see below. It is natural to have some doping dependence 
of the lattice dielectric constant since the lattice dynamics may change 
locally (get softer) with La $\to$ Sr substitution. In our  
dimensionless units, the effective charge $Q$ is
\begin{eqnarray}
Q=\frac{e^2}{\epsilon_l a_0 J}\approx 30/\epsilon_l\;.
\label{effective_charge}
\end{eqnarray} 

According to the same logic the Coulomb interaction between holons 
(\ref{Hhh}) is replaced by
\begin{eqnarray}
\label{uhhL}
U_{ij}&\to&Q\left(
\frac{1}{\sqrt{|{\bf r}_{i}-{\bf r}_{j}|^{2}+a_{HF}^{2}}}\right.
\nonumber\\
&&\left.+\sum_{n=1}^{\infty}\frac{2(-1)^n}
{\sqrt{|{\bf r}_{i}-{\bf r}_{j}|^{2}+(2nd)^{2}}}\right) \;.
\end{eqnarray}

\begin{figure}[htb]
\centering
\leavevmode
\includegraphics[width=0.8\columnwidth,clip=true]{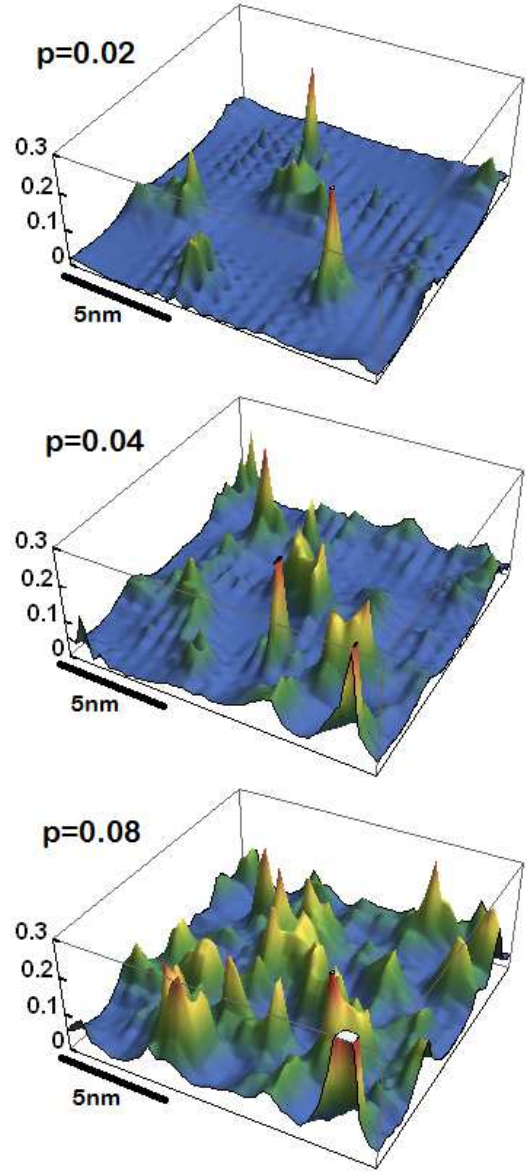}
\caption{The charge density of mobile holes in La$_{2-x}$Sr$_x$CuO$_4$ 
calculated at zero temperature and different doping levels.}
\label{fig:LSCO_density}
\end{figure}

\begin{figure}[htb]
\centering
\leavevmode
\includegraphics[width=0.95\columnwidth,clip=true]{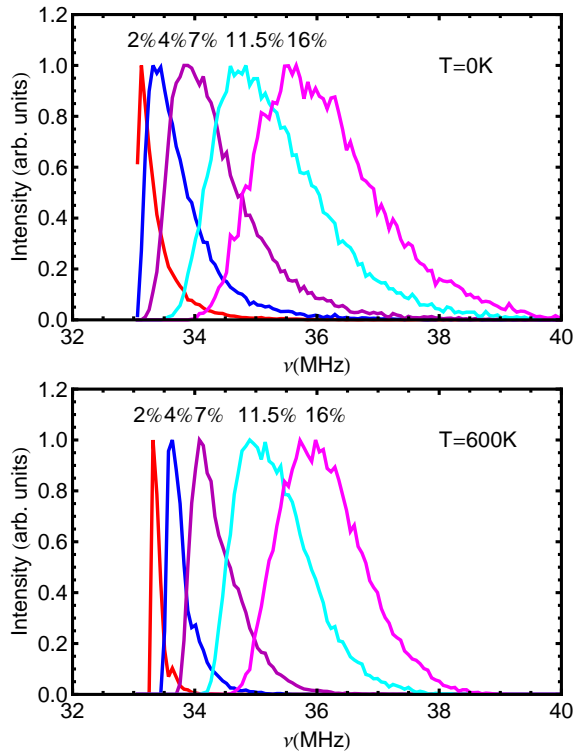}
\caption{(Color online) Theoretical NQR spectra for in-plane $^{63}$Cu
in La$_{2-x}$Sr$_x$CuO$_4$ at $T=0$ and $T=600$~K and for different
doping levels $p$.}
\label{fig:LSCO_NQR}
\end{figure}

The HF Hamiltonian is diagonalized in a $36\times 36$ cluster. The resulting 
charge densities for particular realizations of Coulomb defects at different 
doping levels are shown in Fig.~\ref{fig:LSCO_density}. 
To calculate the NQR spectrum of in-plane $^{63}$Cu, we first calculate 
the hole density distribution averaged over 20 random impurity configurations,
and then convert the hole density at a site $i$ to the NQR frequency 
using the following scaling~\cite{Haase04}
\begin{eqnarray}
\label{sl19}
\nu_{i}=33+19n_{i} \; ({\rm MHz})\;,  
\end{eqnarray}
which implies that the NQR spectrum is directly related to the charge 
density distribution. The resulting NQR spectra for zero temperature and 
for $T=600$~K are presented in Fig.~\ref{fig:LSCO_NQR}.
The theoretical NQR spectra at $T=600$~K agree very well with experimental 
data~\cite{Singer02}. Although we are not aware of NQR data at low 
temperatures, the theoretical spectra at $T=0$ are shown in
order to demonstrate how the charge density distribution evolves with
temperature. It is instructive to present the widths of the charge density
distribution. Our simulation yields 
\begin{eqnarray}
\label{GLSCO}
&&LSCO \ at \ p=0.16 \!\!\ : \ 
\! \left\{
\begin{array}{l}
\Gamma_p(T\!=0)=0.12\\
\Gamma_p(T\!=600K)=0.09,  
\end{array}\right.\nonumber\\
&&LSCO \ at \ p=0.08 \!\!\ : \
\! \left\{
\begin{array}{l}
\Gamma_p(T\!=0)=0.07\\
\Gamma_p(T\!=600K)=0.05. 
\end{array}\right.
\end{eqnarray}

In our previous work,~\cite{Chen09} we performed similar calculations for 
La$_{2-x}$Sr$_x$CuO$_4$ and demonstrated that the Hartree-Fock calculation 
reproduces reasonably well the evolution of NQR lineshapes with doping. 
However, the calculated NQR linewidths were about 20-30\% larger compared to 
the experimental values. The goal of the present calculation is to fit the 
NQR linewidths quantitatively, using the dielectric constant 
(\ref{epsilon_LSCO}) as a fitting parameter. Based on this fit we accurately 
quantify the charge inhomogeneity and the hole mean free path as discussed 
below. We also note that the previous calculation~\cite{Chen09} has found an  
additional high frequency hump/shoulder in the NQR line. The hump was due to 
the strong Coulomb binding of holes to the accidentally clustered Coulomb 
defects. The binding is very sensitive to the strength of the Coulomb 
attraction. This binding does not show up in the present calculation because 
of the larger value of the dielectric constant used, and also due to the fact 
that we impose here the ''nonadjacent'' condition for Coulomb defects [see 
the paragraph before Eq.~(\ref{scr})]. As suggested in 
Ref.~\onlinecite{Singer02}, the experimentally observed high frequency hump 
(the NQR "B-line") is most probably due to the direct action of Sr ion 
Coulomb potential on the Cu nucleus, which is beyond the scope of the 
present model.

\subsection{HgBa$_2$CuO$_{4+\delta}$}

\begin{figure}[htb]
\centering
\leavevmode
\includegraphics[width=0.75\columnwidth,clip=true]{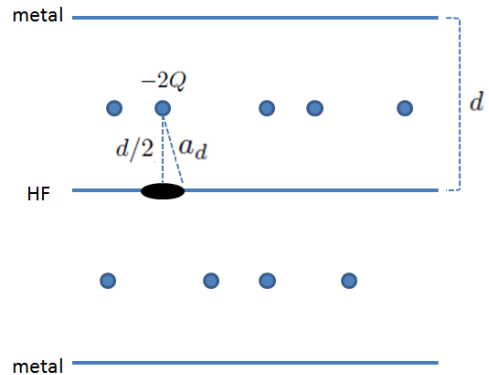}
\caption{(Color online) The Hartree-Fock model of HgBa$_2$CuO$_{4+\delta}$ 
with two layers of Coulomb defects (interstitial oxygen ions) and two 
''metallic'' sheets.}
\label{fig:HBCO_structure}
\end{figure}

Interstitial oxygen ions in HgBa$_2$CuO$_{4+\delta}$ are located right between 
two neighboring CuO$_{2}$ layers, on top of four adjacent Cu 
sites.~\cite{Gippius97} The minimal model for HgBa$_2$CuO$_{4+\delta}$ is 
shown in Fig.~\ref{fig:HBCO_structure}. 
There is one layer of oxygen defects above the CuO$_{2}$ plane and another 
defect layer below it. Concentration of defects in each defect layer 
is $\delta$. We assume that each interstitial oxygen donates two holes into 
the CuO$_2$ plane, i.e., 
\begin{equation}
p=2\delta \;.
\end{equation}
According to this picture, the holon interaction with a defect, Eq.~(\ref{HUQ})
is replaced by
\begin{eqnarray}
{\cal U}_{li}&\to&-2Q\left(
\frac{1}{\sqrt{|{\bf R}_{l}-{\bf r}_{i}|^2 + a_{d}^2}}\right. \\
&&+\sum_{n=1}^{\infty}\frac{(-1)^n}
{\sqrt{|{\bf R}_{l}-{\bf r}_{i}|^{2}+(2nd+d/2)^{2}}}
\nonumber \\
&&\left.+\sum_{n=1}^{\infty}\frac{(-1)^n}
{\sqrt{|{\bf R}_{l}-{\bf r}_{i}|^{2}+(2nd-d/2)^{2}}}\right) \;. 
\nonumber
\end{eqnarray}
The interlayer distance in HgBa$_2$CuO$_{4+\delta}$ is $d=2.5$, and each 
interstitial oxygen carries ''charge'' $-2Q$. Taking into account size of 
the Zhang-Rice singlet, we have $a_{d}=\sqrt{a_{ZR}^{2}+(d/2)^{2}}\approx 1.5$. 
The hole-hole interaction has exactly the same form as that in 
La$_{2-x}$Sr$_x$CuO$_4$, Eq.~(\ref{uhhL}), with the interlayer distance
$d=2.5$ and with the value of $Q$ described below. In each defect layer we 
simulate positions of the defects randomly, and impose again the nonadjacent 
condition for the defects. 

Similar to the procedure in La$_{2-x}$Sr$_x$CuO$_4$, we fine tune the NQR 
widths using the lattice dielectric constant $\epsilon_l$ as fitting 
parameter. We find that in order to fit experimental NQR 
spectra~\cite{Rybicki09,Gippius97} the dielectric constant has to be taken as 
\begin{eqnarray}
\epsilon_l=30(1+25p)\;.
\label{epsilon_HBCO}
\end{eqnarray}
Note that the doping dependence of $\epsilon_l$ in this case is four times 
stronger than that in Eq.~(\ref{epsilon_LSCO}). The stronger dependence is 
quite natural since it is due to the shift of the interstitial oxygen position 
in an applied electric field. The shift is significant because binding of 
interstitial oxygens in the lattice is weak. In other words, a dopant oxygen
brings in new local phonon modes which enhance the dielectric constant. 
The effective charge $Q$ is determined by the same 
Eq.~(\ref{effective_charge}), with $\epsilon_l$ from Eq.~(\ref{epsilon_HBCO}).
 
\begin{figure}[htb]
\centering
\leavevmode
\includegraphics[width=0.8\columnwidth,clip=true]{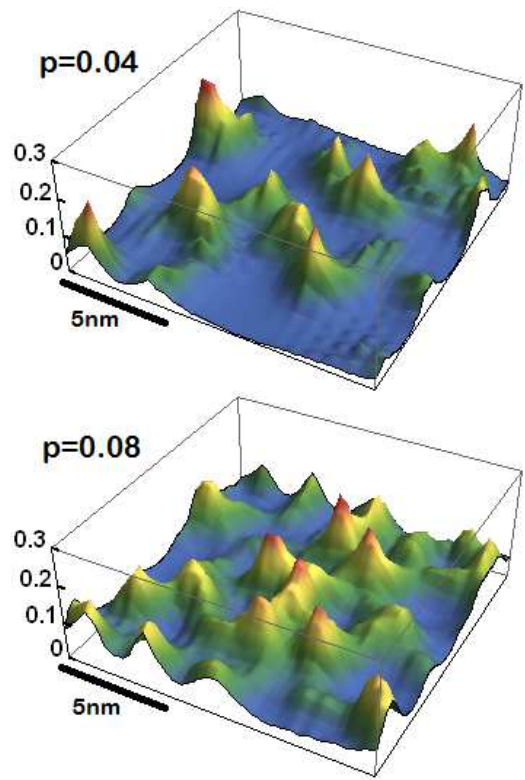}
\caption{The charge density of mobile holes in HgBa$_2$CuO$_{4+\delta}$
at zero temperature and at different doping levels.}
\label{fig:HBCO_density}
\end{figure}
\begin{figure}[htb]
\centering
\leavevmode
\includegraphics[width=0.80\columnwidth,clip=true]{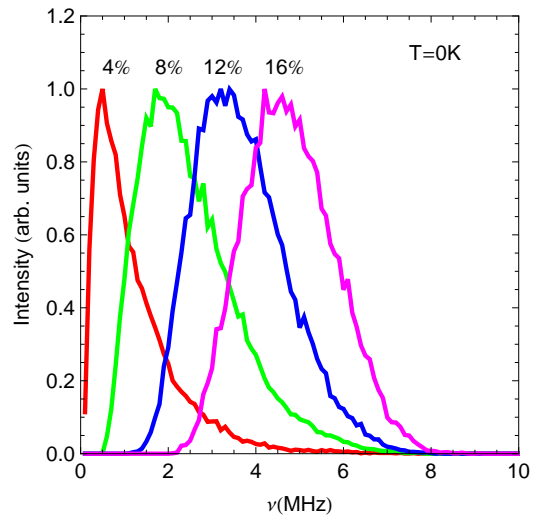}
\caption{(Color online) Theoretical NQR spectra for in-plane
$^{63}$Cu in HgBa$_2$CuO$_{4+\delta}$ at zero temperature and for
different doping levels $p$.}
\label{fig:HBCO_NQR}
\end{figure}
The HF Hamiltonian is diagonalized in a $36\times 36$ cluster. The resulting 
charge densities for particular realizations of Coulomb defects at different 
doping levels are shown in Fig.~\ref{fig:HBCO_density}. A strong charge 
inhomogeneity induced by oxygen dopants is apparent. However, density 
distribution profiles here are much smoother than in La$_{2-x}$Sr$_x$CuO$_4$ 
as one can easily see by comparing Figs.~\ref{fig:HBCO_density} and 
\ref{fig:LSCO_density} at same doping levels.

Now, we address the NQR spectrum of the in-plane $^{63}$Cu in 
HgBa$_2$CuO$_{4+\delta}$. We first calculate the hole density distribution 
averaged over 20 random impurity configurations, and then convert the hole 
density at a site $i$ to the NQR frequency by
\begin{eqnarray}
\label{sl30}
\nu_{i}=C+30n_{i} \; ({\rm MHz})\;, 
\end{eqnarray}
where $C$ is a doping independent constant. Note that the coefficient 30 MHz 
in this formula is different from that in Eq.~(\ref{sl19}). We found the 
coefficient in Eq.~(\ref{sl30}) by fitting the NQR-line centers using the 
experimental data of Ref.~\onlinecite{Gippius97} (this way is more reliable 
than using results of theoretical calculations of electric field 
gradients.~\cite{Ambrosch-Draxl06}) Since NQR data for HgBa$_2$CuO$_{4+\delta}$ 
are taken at sufficiently low temperatures, we perform calculations only at 
$T=0$. Fig.~\ref{fig:HBCO_NQR} shows the theoretical NQR lines by assuming 
$C=0$, since it is an irrelevant constant shift. The calculations agree well 
with the experimental data of Refs.~\onlinecite{Gippius97,Rybicki09}. At 
optimal doping $p=0.16$, the NQR linewidth in HgBa$_2$CuO$_{4+\delta}$ is 
$\Gamma_{NQR}\approx 2.6$~MHz. The widths of the charge density distribution 
are 
\begin{eqnarray}
\label{GHBCO}
&&HBCO \  at \ p=0.16 \ : \ \Gamma_p(T=0)=0.09 \ , \nonumber\\
&&HBCO \  at \ p=0.08 \ : \ \Gamma_p(T=0)=0.08 \ .
\end{eqnarray}
Comparing with the corresponding values for La$_{2-x}$Sr$_x$CuO$_4$ in 
Eq.~(\ref{GLSCO}), we see that at zero temperature the amplitudes of charge 
inhomogeneities in HgBa$_2$CuO$_{4+\delta}$ and in La$_{2-x}$Sr$_x$CuO$_4$ are 
quite similar. At optimal doping $p=0.16$, the amplitude of charge 
inhomogeneity in HgBa$_2$CuO$_{4+\delta}$ is only by 30\% smaller than that in
La$_{2-x}$Sr$_x$CuO$_4$.

\subsection{Density of states and mean free path}

\begin{figure}[h!]
\centering
\leavevmode
\includegraphics[width=0.79\columnwidth,clip=true]{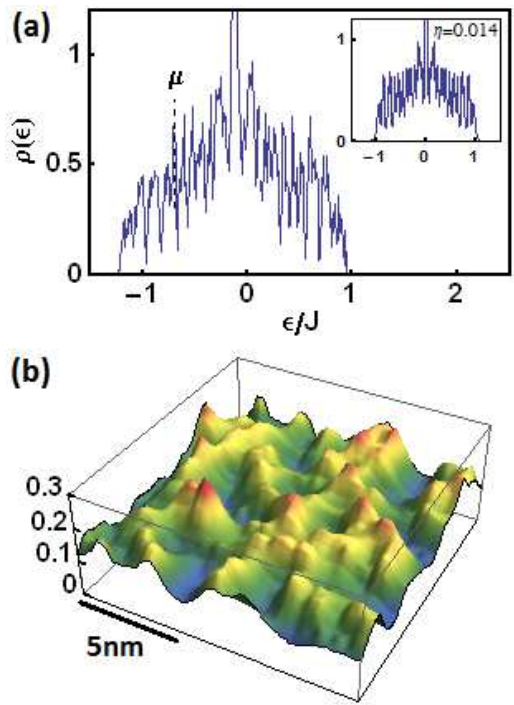}
\caption{(Color online) (a) Density of states and (b) charge density 
profile in HgBa$_2$CuO$_{4+\delta}$ at optimal doping $p=0.16$. Shown in the 
inset of panel (a) is the DOS of the homogeneous model with effective 
scattering rate $\eta=0.014$.}
\label{fig:comparison_HBCO}
\end{figure}

\begin{figure}[h!]
\centering
\leavevmode
\includegraphics[width=0.79\columnwidth,clip=true]{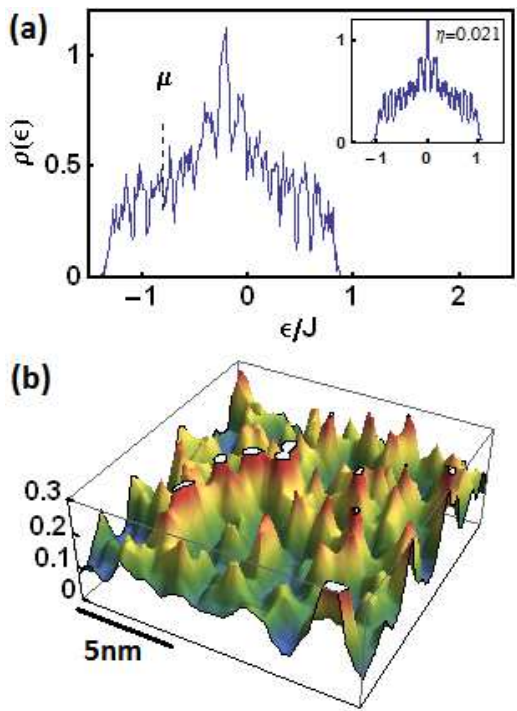}
\caption{
(Color online)(a) DOS and (b) charge density profile 
in La$_{2-x}$Sr$_x$CuO$_4$ at $p=0.16$. The inset shows the DOS 
of the homogeneous model with scattering rate $\eta=0.021$.}
\label{fig:comparison_LSCO}
\end{figure}

A comparison between Eqs.~(\ref{GLSCO}) and (\ref{GHBCO}) suggests that the 
hole density inhomogeneity in HgBa$_{2}$CuO$_{4+\delta}$ is pretty close to 
that in La$_{2-x}$Sr$_x$CuO$_4$. This seems rather surprising, given that the 
superconducting critical temperature in HgBa$_2$CuO$_{4+\delta}$ is much
higher than that in La$_{2-x}$Sr$_x$CuO$_4$. Indeed, a disorder reduces the 
$d-$wave superconducting critical temperature and, based on this argument, 
one can expect that La$_{2-x}$Sr$_x$CuO$_4$ is ''more disordered''. However, 
one should quantify exactly what is the measure of the disorder. While the 
overall amplitude of the charge inhomogeneity is one possible factor, the 
hole mean free path is another and in fact more important measure of the 
disorder. We already noticed that the spatial charge distribution profiles 
induced by interstitial oxygens and Sr-dopants are rather different. To 
quantify this difference in more detail, we calculate the quasiparticle DOS 
defined in the standard way
\begin{eqnarray}
\label{ro}
\rho(\epsilon)=\frac{1}{N}\sum_{n}\delta(\epsilon-E_{n})\;, 
\end{eqnarray}
where $N$ is the total number of sites, and $E_{n}$ is the $n$-th eigenenergy.  
Plots of the DOS in HgBa$_2$CuO$_{4+\delta}$ and in La$_{2-x}$Sr$_x$CuO$_4$ at 
doping $p=0.16$ are shown in Figs.~\ref{fig:comparison_HBCO} and 
\ref{fig:comparison_LSCO}, together with the charge profiles for particular 
disorder realizations. 
The DOS has been obtained after averaging over 40 different disorder 
realizations. We have checked that the DOS calculated with 20 realizations 
of random dopant positions is practically the same as the DOS calculated 
with 40 realizations. This means that 40 realizations is sufficient to 
disregard the statistical noise. The calculated DOS exhibits pronounced 
oscillations. These oscillations are a byproduct of the finite size of the 
cluster. Maxima of the DOS correspond to degenerate states with dispersion 
(\ref{ht1}) on the 36$\times$36 torus. The oscillations must certainly 
disappear in the thermodynamic limit. However, oscillations have a physical 
meaning: they indicate that the quantum states are quite extended, with the 
hole mean free path $l$ comparable with the size of the cluster used. To 
estimate the mean free path more accurately we use the following procedure. 
We consider the DOS of an ideal homogeneous system described by the Hamiltonian 
(\ref{ht}). It consists of $\delta$-functions whose positions are fixed by 
periodic boundary conditions. The weight of each $\delta$-function can be 
easily calculated for the 36$\times$36 torus. Now, we artificially broaden 
each $\delta$-function, 
\begin{equation}
\delta(\epsilon-\epsilon_n)\to \frac{1}{\pi}
\frac{\eta}{(\epsilon-\epsilon_n)^2 +\eta^2} \ ,
\end{equation}
to simulate a disorder scattering. We find that the DOS of this model 
is very sensitive to $\eta$. Then, we adjust the broadening $\eta$ 
to reproduce (roughly) the amplitudes of the DOS oscillations obtained in our 
actual calculations, as shown in Figs.~\ref{fig:comparison_HBCO}(a) and 
\ref{fig:comparison_LSCO}(a). This gives the following values of the 
effective broadening: $\eta=0.014$ for HgBa$_2$CuO$_{4.08}$, and $\eta=0.021$ 
for La$_{1.84}$Sr$_{0.16}$CuO$_4$. The mean free path of a hole at the Fermi 
surface can be then estimated as: 
\begin{eqnarray}
l=v_{F}\tau=\frac{v_{F}}{2\eta} \ ,
\end{eqnarray}
where $\tau=1/2\eta$ is the collision time. At small doping, $p \ll 1$, the 
dispersion (\ref{ht1}) results in the Fermi velocity 
$v_{F}\approx 8t^{\prime\prime}a_0\sqrt{\pi p}$. Hence, at $p=0.16$, the Fermi 
velocity is $v_{F}\approx 1.4Ja_0$. Together with the above values of $\eta$, 
this results in the following estimates for the hole mean free paths: 
\begin{eqnarray}
\label{ml}
&&HgBa_2CuO_{4.08}: \ \ \ \ \ l\approx 50a_0\approx 19nm \ , \nonumber\\
&&La_{1.84}Sr_{0.16}CuO_4: \ \ l\approx 34a_0\approx 13nm \ . 
\end{eqnarray}
We thus find that La$_{2-x}$Sr$_x$CuO$_4$ is indeed more disordered, in the 
sense that it has a shorter mean free path. The different mean free paths 
are due to different distances from the CuO$_2$ plane to the Coulomb 
defect (interstitial oxygen versus Sr-dopant), and due to different 
distances between Coulomb defects. In HgBa$_2$CuO$_{4+\delta}$, both 
distances are larger, therefore the Coulomb potential is smoother and hence 
less contributes to the large angle scattering. This difference is clearly 
reflected in the charge density profiles shown in 
Fig.~\ref{fig:comparison_HBCO}(b) and \ref{fig:comparison_LSCO}(b).
Overall amplitudes of the charge inhomogeneity are pretty close, but the 
charge distribution in La$_{2-x}$Sr$_x$CuO$_4$ is much more spiky. This gives 
rise to the strong scattering with large momentum transfer which is detrimental 
for $d-$wave superconductivity.

Considering the superconducting correlation length $\xi \approx 2$~nm, one
finds that $l/\xi \sim 6$ in La$_{2-x}$Sr$_x$CuO$_4$ and $l/\xi \sim 10$ in 
HgBa$_2$CuO$_{4+\delta}$. One should therefore expect that the disorder 
suppression of superconductivity in HgBa$_2$CuO$_{4+\delta}$ is indeed
weaker. In addition, according to our fits of the NQR data, the effective ionic 
dielectric constant $\epsilon_l$ in HgBa$_2$CuO$_{4+\delta}$, 
Eq.~(\ref{epsilon_HBCO}), is larger than that in La$_{2-x}$Sr$_x$CuO$_4$, 
Eq.~(\ref{epsilon_LSCO}). At the optimal doping $p=0.16$, for example, 
$\epsilon_l(LSCO)=60$ and $\epsilon_l(HBCO)=150$. The larger $\epsilon_l$ 
in HgBa$_2$CuO$_{4+\delta}$ implies better screening of the Coulomb repulsion 
between holes and hence a smaller Coulomb pseudopotential. This may further 
enhance $T_c$ of HgBa$_2$CuO$_{4+\delta}$ compared to that of 
La$_{2-x}$Sr$_x$CuO$_4$.

\section{Nanoscale Hole Density Inhomogeneity in the surface layer 
of $\mbox{Bi}_2\mbox{Sr}_2\mbox{CaCu}_2\mbox{O}_{8+\delta}$}

Recent STM experiments~\cite{McElroy05,Gomes07,Pasupathy08} have revealed
large variations of pairing gaps in Bi$_2$Sr$_2$CaCu$_2$O$_{8+\delta}$ which
are spatially correlated with the dopant oxygen density. Motivated by this 
observation, we now consider the charge distribution in the surface CuO$_2$ 
layer of this compound. Our purpose is to estimate the magnitude of the charge 
inhomogeneity induced by interstitial oxygen dopants, and see how the hole 
density profile is correlated with the position of these oxygens. In the 
present work, we do not calculate the local DOS and hence cannot address 
STM directly. Instead, we calculate the charge distribution similar to the 
previous sections, and show that the effect of interstitial oxygen on charge 
inhomogeneity is very significant. 

As we have seen above, charge distribution profiles depend on the lattice 
dielectric constant whose values can be reliably obtained by fitting the 
NQR data. Due to lack of systematic NQR data, we cannot determine the 
dielectric constant of Bi$_2$Sr$_2$CaCu$_2$O$_{8+\delta}$ in this way. 
Instead, we assume that the doping dependence of the lattice dielectric 
constant in the bulk of Bi$_2$Sr$_2$CaCu$_2$O$_{8+\delta}$ is described by 
the same formula Eq.~(\ref{epsilon_HBCO}) as in HgBa$_2$CuO$_{4+\delta}$. 
This is because both compounds are doped by interstitial oxygens which have 
similar influence on lattice dynamics and dielectric screening. This adoption 
certainly ignores lattice structure details and should be further refined 
when NQR data is available. Nevertheless, we found that our results for  
Bi$_2$Sr$_2$CaCu$_2$O$_{8+\delta}$ are fairly robust within a sensible 
variations of the dielectric constant, and thus they should give 
a qualitative description of the surface charge inhomogeneity.

\begin{figure}[htb]
\centering
\leavevmode
\includegraphics[width=0.75\columnwidth,clip=true]{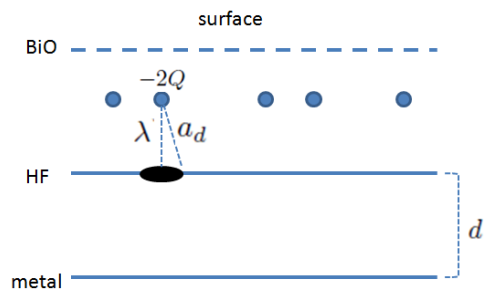}
\caption{(Color online) The Hartree-Fock model for 
Bi$_2$Sr$_2$CaCu$_2$O$_{8+\delta}$ surface. Blue dots indicate the
interstitial oxygen dopants.}
\label{fig:BSCCO_structure}
\end{figure}
Bi$_2$Sr$_2$CaCu$_2$O$_{8+\delta}$ is a double layer compound and a schematic 
picture of the two surface CuO$_2$ layers is shown in 
Fig.~\ref{fig:BSCCO_structure}. Interstitial oxygens are assumed to be 
located at $\lambda\approx 0.37$~nm above the CuO$_2$ layer, which is shown to 
be the most energetically favorable position.~\cite{He} We assume that the 
surface concentration of Coulomb defects (interstitial oxygens) is $\delta$. 
They are randomly distributed and, similarly to the previous considerations, 
we impose again the condition that the defects never sit next to each other, 
i.e., the distance between the defects is always larger than $\sqrt{2}$ 
(in units of $a_0$). The separation between the CuO$_2$ layers in the double 
layer structure is $d=0.33$~nm$\approx 0.86$. We treat the top layer by  
Hartree-Fock approximation, and the underneath layer as a ''metallic'' sheet 
that provides screening. There are also Coulomb defects underneath of the 
screening layer, but they are well screened by the metallic sheet and hence 
do not influence the Hartree-Fock procedure. The hole-defect and the 
hole-hole interaction in the Hartree-Fock top layer are of the following form: 
\begin{eqnarray}
{\cal U}_{li}&\to&-2Q\left(
\frac{1}{\sqrt{|{\bf R}_{l}-{\bf r}_{i}|^{2}+a_d
^{2}}}\right.
\nonumber \\
&&\left.-\frac{1}{\sqrt{|{\bf R}_{l}-{\bf r}_{i}|^{2}+(2d+\lambda)
^{2}}}\right)\;,
\nonumber \\
U_{ij}&\to&Q\left(
\frac{1}{\sqrt{|{\bf r}_{i}-{\bf r}_{j}|^{2}+a_{HF}^{2}}}\right.
\nonumber \\
&&\left.-\frac{1}{\sqrt{|{\bf r}_{i}-{\bf r}_{j}|^{2}+(2d)^{2}}}\right)\;.
\end{eqnarray}
Here $a_{d}=\sqrt{\lambda^{2}+a_{ZR}^{2}}\approx 1.28$,
and $a_{HF}\approx\sqrt{2a_{ZR}^{2}}\approx 1$. Note that there is only one
image charge per physical charge because there is only one screening
layer. We assume that the lattice contribution to the dielectric constant in 
Bi$_2$Sr$_2$CaCu$_2$O$_{8+\delta}$ has the form Eq.~(\ref{epsilon_HBCO}). 
On the surface, the dielectric constant is expected to be reduced to half 
of its value in the bulk,~\cite{BT} therefore 
\begin{eqnarray}
\epsilon^{\rm BSCCO}=15(1+25p)\;, 
\label{epsilon_BSCCO}
\end{eqnarray} 
and the value of effective ''charge'' $Q$ follows from 
Eq.~(\ref{effective_charge}). 

\begin{figure}[h!]
\centering
\leavevmode
\includegraphics[width=0.95\columnwidth,clip=true]{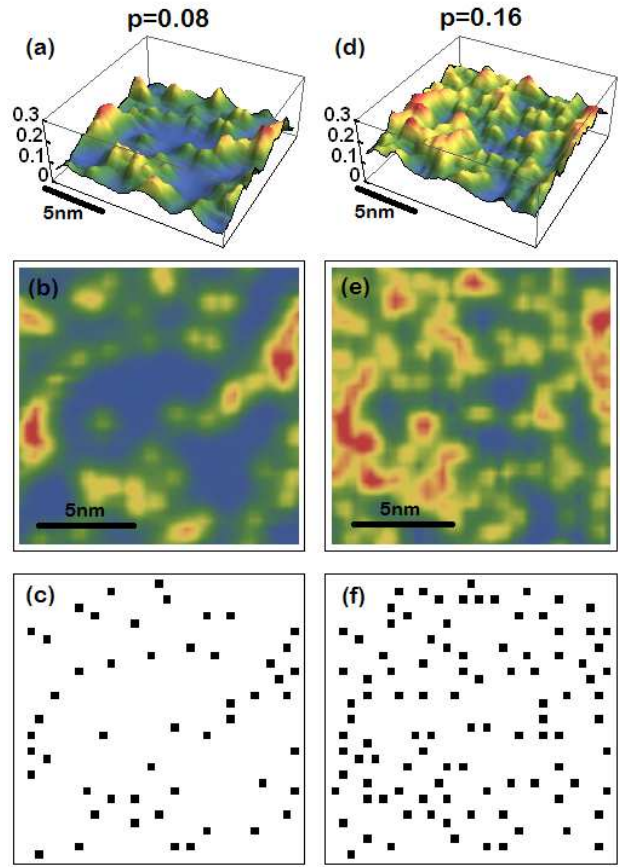}
\caption{(Color online) (a,d) The zero temperature hole density plots and 
(b,e) the hole density maps for the surface layer of 
Bi$_2$Sr$_2$CaCu$_2$O$_{8+\delta}$ for the average hole density    
$p=0.08$ (left panels) and $0.16$ (right panels). The lower panels (c) 
and (f) show positions of the oxygen dopants for the corresponding 
realizations of disorder. It is evident that the dopant oxygens locally 
increase the hole density.}
\label{fig:BSCCO_density}
\end{figure}
\begin{figure}[h!]
\centering
\leavevmode
\includegraphics[width=0.95\columnwidth,clip=true]
{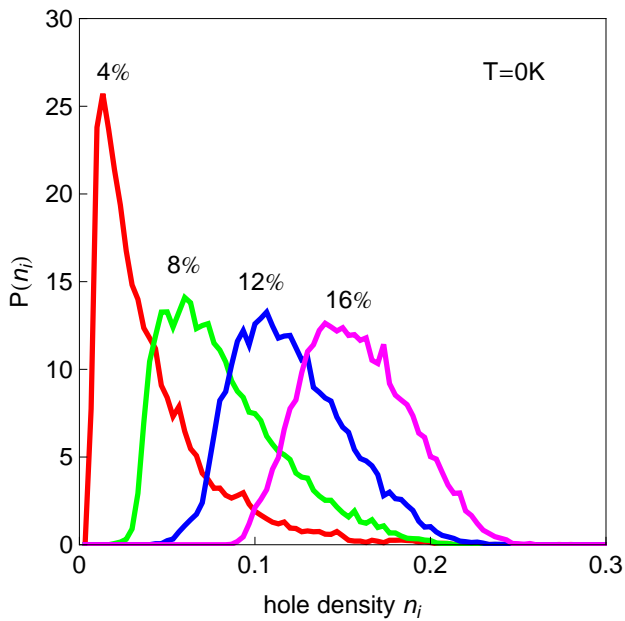}
\caption{(Color online) The zero temperature hole density distribution in 
the surface layer of Bi$_2$Sr$_2$CaCu$_2$O$_{8+\delta}$ at various doping 
levels $p$. Each curve is averaged over 20 realizations of disorder.}
\label{fig:BSCCO_density_distribution}
\end{figure}
Plots of the in-plane hole density for $\delta=0.04$ ($p=0.08$) and 
$\delta=0.08$ ($p=0.16$) are presented in Fig.~\ref{fig:BSCCO_density}, 
together with corresponding maps for these particular realizations of 
Coulomb defects. The plots demonstrate a large spatial variations in hole 
density. For the $\delta=0.04$ case, the range of hole density modulation is 
about $0.03<n_{i}<0.15$, which is a very significant fluctuation, having in 
mind that the average hole density is $\langle n_{i}\rangle =p=0.08$. The same 
strong inhomogeneity is seen in the $\delta=0.08$ case, where the local density 
varies in the range roughly $0.10<n_{i}<0.23$, while average density is  
$\langle n_{i}\rangle =p=0.16$. The density distribution curves for different 
doping levels are shown in Fig.~\ref{fig:BSCCO_density_distribution}. We 
stress that the precise density profiles depend on the dielectric constant 
which, due to lack of NQR data for Bi$_2$Sr$_2$CaCu$_2$O$_{8+\delta}$, is 
taken here in an {\it ad hoc} way. On a qualitative level, however, the 
density inhomogeneity is rather stable with respect to the value of the 
dielectric constant. For example, the widths of the surface density 
distributions plotted in Fig.~\ref{fig:BSCCO_density_distribution} are only 
slightly larger than those obtained in the case of HgBa$_2$CuO$_{4+\delta}$, see 
Eq.~(\ref{GHBCO}). This is in spite of the fact that the results for 
HgBa$_2$CuO$_{4+\delta}$ have been obtained at the twice larger value of the 
dielectric constant. Therefore, we believe that our conclusion about the 
strong surface charge density inhomogeneity is very reliable.

\begin{figure}[htb]
\centering
\leavevmode
\includegraphics[width=0.95\columnwidth,clip=true]
{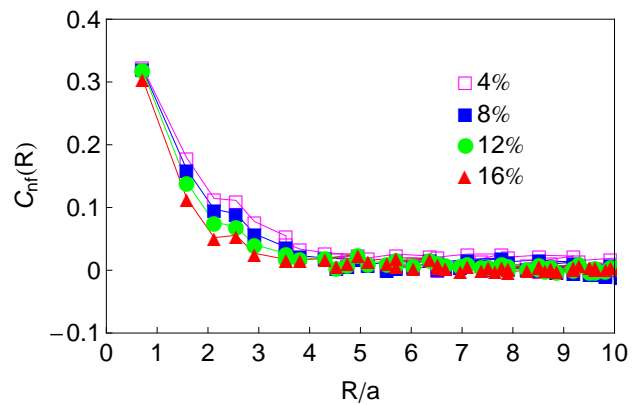}
\caption{(Color online) The correlation function Eq.~(\ref{cc}) between the 
interstitial oxygen position and the local density of mobile holes for 
different values of doping $p$.}
\label{fig:BSCCO_density_oxygen_correlator}
\end{figure}

Naturally, the charge density of mobile holes is higher around areas with 
higher interstitial oxygen concentration. Fig.~\ref{fig:BSCCO_density} 
clearly shows this correlation. To quantify this, we define 
a correlation function between local hole density and dopant oxygen 
positions, which is analogous to the correlator introduced in 
Ref.~\onlinecite{McElroy05} for analysis of the spatial variations of the 
local DOS. On a discrete lattice, the hole density $n_{j}$ is defined on 
sites of the square lattice representing the CuO$_2$ plane. The function 
$f_{i}$ indicates the location of interstitial oxygens: 
\begin{eqnarray}
f_{i}=\left\{\begin{array}{ll} 1 & {\rm if\;{\it i}\in dopant \; oxygen\;,} 
\\ 0 & {\rm elsewhere \; ,} \end{array} \right. 
\end{eqnarray}
where $i$ runs through points at the center of the plaquettes.
The density-oxygen correlation function is then defined as 
\begin{eqnarray}
\label{cc}
C_{nf}({\bf R})=
\frac{1}{N}\sum_{i}\frac{\left[f_{i}-\overline{f}\right]\left[n_{i+{\bf R}}
-\overline{n}\right]}{\sqrt{A_{f}A_{n}}}\;,
\end{eqnarray}
with proper normalizations
\begin{eqnarray}
A_{f}&=&\frac{1}{N}\sum_{i}\left(f_{i}-\overline{f}\right)^{2}\;,
\nonumber \\
A_{n}&=&\frac{1}{N}\sum_{j}\left(n_{j}-\overline{n}\right)^{2}\ ,
\end{eqnarray} 
where $\overline{f}=\frac{1}{N}\sum_{i}f_{i}=\delta$ and
$\overline{n}=\frac{1}{N}\sum_{j}n_{j}=p$.
Fig.~\ref{fig:BSCCO_density_oxygen_correlator} shows the correlation function 
$C_{nf}({\bf R})$ averaged over 20 disorder realizations for each doping.
There is a clear positive correlation due to the Coulomb attraction to the 
oxygen defects. The value of the correlator at 
$R \to 0$ is $C \approx 0.3-0.4$ and the scale at which it goes to zero is  
about 10-15~\AA. Interestingly, the correlator between the gap in local DOS 
and the interstitial oxygen position measured in STM shows the same positive 
correlation with very similar scales~\cite{McElroy05}. Further investigation 
is necessary to clarify if there is a connection between these two 
correlators, and to understand the physical reasons behind this apparent 
correspondence. 

\section{Conclusions}

In this paper, we study the spatial distribution of doped holes in cuprates,
focusing particularly on a comparison between two different physical
situations: doping by a cationic substitution Sr for La as in 
La$_{2-x}$Sr$_x$CuO$_4$, and doping by interstitial oxygen ions as in       
HgBa$_2$CuO$_{4+\delta}$. The main results are summarized as follows.

The hole density inhomogeneity in HgBa$_2$CuO$_{4+\delta}$ is nearly as strong 
as in La$_{2-x}$Sr$_x$CuO$_4$. For example, at the optimal doping $p=0.16$ the 
width of the charge density distribution is about $\Gamma_p=0.09$, which is 
close to the corresponding number $\Gamma_p=0.12$ in La$_{2-x}$Sr$_x$CuO$_4$. 
This conclusion is well supported by the comparison of our calculations
with the existing NQR data in HgBa$_2$CuO$_{4+\delta}$ and 
La$_{2-x}$Sr$_x$CuO$_4$. In spite of the close overall amplitudes, the
landscape of charge inhomogeneity in these two compounds are very different. 
In oxygen doped HgBa$_2$CuO$_{4+\delta}$, the disorder potential profiles 
are much smoother than that in Sr-doped La$_{2-x}$Sr$_x$CuO$_4$. 
Correspondingly, the hole mean free path in HgBa$_2$CuO$_{4+\delta}$ is 
larger. In other words, disorder induced scattering processes with a large 
momentum transfer (which are destructive for $d-$wave pairing) are less 
pronounced in oxygen doped HgBa$_2$CuO$_{4+\delta}$ compared to the case of 
La$_{2-x}$Sr$_x$CuO$_4$. In addition, the screening of the Coulomb repulsion 
between holes in HgBa$_2$CuO$_{4+\delta}$ is about twice stronger than that 
in La$_{2-x}$Sr$_x$CuO$_4$. In our opinion, these two reasons might explain 
the much higher superconducting critical temperature of 
oxygen doped HgBa$_2$CuO$_{4+\delta}$. 

We found that the charge density nanoscale inhomogeneity in the surface CuO$_2$ 
layer of Bi$_2$Sr$_2$CaCu$_2$O$_{8+\delta}$ (the layer available for STM) is 
of the same magnitude as that in the bulk of HgBa$_2$CuO$_{4+\delta}$. As
expected on physical grounds, the hole density positively correlates with 
the positions of interstitial dopant oxygen. Remarkably, the correlation 
function obtained here resembles the positive correlation between the local 
gap and dopant oxygens seen in the STM data. The reason for this apparent 
coincidence and implications of the charge inhomogeneity for the spatial 
variations of the pairing gaps in Bi$_2$Sr$_2$CaCu$_2$O$_{8+\delta}$ have 
to be clarified in future studies. 

\section{Acknowledgments}
 
We would like to thank J. Haase for stimulating discussions of the NQR data 
in cuprates, and P.J. Hirschfeld, H. Alloul, M.-H. Julien, E.Ya. Sherman 
for useful communications. G.Kh. thanks the School of Physics and the Gordon 
Godfrey fund, UNSW, for kind hospitality. Numerical calculations are done 
by using facilities of Australian National Computational Infrastructure.


\begin{thebibliography}{99}
\bibitem{com1} In principle, it is possible to fit the NQR observed charge 
inhomogeneity even with the large Fermi surface. However, in this case
one must significantly reduce the lattice dielectric constant $\epsilon_l$.
In particular, one has to take $\epsilon_l \approx 5-6$ instead of
the experimentally known values $\epsilon_l \approx 30-40$.
\bibitem{noteMBZ} Note that usage of the magnetic Brillouin zone notations 
in the $t$-$J$ model does not imply that there is a long range magnetic order. 
This is just a convenient way to avoid double counting of degrees of freedom 
related to the pseudospin, see discussion in Ref.~\onlinecite{CST}.
\end{thebibliography}
\end{document}